# Effect of Zeeman splitting on magnetoresistivity of 2D hole gas in a Ge$_{1-x}$Si$_x$/Ge/Ge$_{1-x}$Si$_x$ quantum well

Yu.G. Arapov[1], V.N. Neverov[1], G.I. Harus[1], N.G. Shelushinina[1], M.V. Yakunin[1],
O.A. Kuznetsov[2], A. deVisser[3], and L. Ponomarenko[3]

[1] *Institute of Metal Physics RAS, Ekaterinburg 620219, Russia*
[2] *Physico-Technical Institute at Nizhnii Novgorod State University, Nizhnii Novgorod, Russia*
[3] *Van der Waals - Zeeman Institute, University of Amsterdam, The Netherlands*

For a two-dimensional (2D) hole system (confined within Ge layers of a multilayered p-Ge/Ge$_{1-x}$Si$_x$ heterostructure) described by Luttinger Hamiltonian with the g-factor highly anisotropic for orientations of magnetic field perpendicular and parallel to the 2D plane ($g_\perp \gg g_\parallel$), reported is an observation of low-temperature transition from metallic (d$\rho$/d$T > 0$) to insulator (d$\rho$/d$T < 0$) behavior of resistivity $\rho(T)$ induced by a perpendicular magnetic field $B$. The revealed positive magnetoresistance scales as a function of $B/T$. We attribute this finding to a suppression of the triplet channel of electron-electron (hole-hole) interaction due to Zeeman splitting in the hole spectrum.

## 1. Introduction

For the diffusive motion of electron in disordered conductors the quantum corrections to the Drude conductivity $\sigma_0$ appear due to both the single-particle weak localization (WL) effects and the disorder modified electron-electron interaction (EEI) [1, 2]. Recently an observation of apparent metal–insulator transition in high mobility semiconductor heterostructures (see the pioneer work [3] and references [4, 5] for an extensive review) has provoked a breakthrough in the theory of EEI effects for two-dimensional (2D) disordered systems [6, 7]. A general theory of the interaction induced quantum corrections to the conductivity tensor of 2D electrons is developed at $kT \ll E_F$ for arbitrary relation between $kT$ and $\hbar/\tau$ (where $\tau$ is the elastic mean free time) in the whole range of temperatures from diffusive ($kT\tau/\hbar \ll 1$) to the ballistic ($kT\tau/\hbar \gg 1$) regimes both for short-range (point-like) [6] and long-range (smooth) [7] random impurity potentials.

According to these latest theories, a linear increase of resistivity $\rho$ with temperature in high-mobility Si-MOSFETs at large values of $\sigma \gg e^2/h$, which for a decade has been considered as a signature of the "anomalous metallic" state, can now be described quantitatively in terms of the interaction effects in the ballistic regime [8]. But the nonmonotonic temperature dependence of $\rho(T)$ near the conjectural conductor – insulator transition (at $\sigma \geq e^2/h$) [8, 9, 10] does not have yet a generally accepted understanding. It is the subject of our investigation, realized on multilayer p-Ge/Ge$_{1-x}$Si$_x$ heterostructures.

Also we investigate a magnetoresistance in a *perpendicular* to the 2D plane magnetic field $B$ where both the Zeeman splitting and WL dephasing effects should be taken into account. We extensively use some ideas exploited for the interpretation of experimental data for $\rho(B, T)$ dependencies of samples with parameters in a vicinity of conjectural conductor – insulator transition in high-mobility 2D semiconductor systems [9, 11-13].

## 2. Experimental results and discussion

Experimental data are presented and analyzed for two similar samples of a multilayered Ge/Ge$_{1-x}$Si$_x$ p-type heterostructure with the number of periods (Ge + GeSi) $N = 15$; the width of quantum wells (Ge layers) $d_w = 80$Å and width of barriers (GeSi layers) $d_b = 120$Å. The central part of each barrier is doped with boron. The hole density and Hall mobility, as obtained from zero field resistivity $\rho_0$ and low field Hall $\rho_{xy}(B)$ at 4.2K, are $p_s = 1.1(1.4) \times 10^{11}$cm$^{-2}$ and

$\mu = 4.0(3.1) \times 10^3$ cm$^2$/Vs [$\rho = 15(16)$ k$\Omega/\square$ and $\varepsilon_F\tau/\hbar = 0.86(0.8)$]. The longitudinal and Hall resistivities have been investigated in magnetic fields $B \leq 5$T at temperatures $T = (0.3 \div 4.2)$K. The data for the two samples are alike, so we shall concentrate on results for one of them.

For these samples, a nonmonotonic temperature behavior of zero-field resistivity is revealed (Fig. 1a), in contrast to a lot of results obtained earlier on samples of the same heterosystem with higher hole densities and mobilities, for which the logarithmic drop of $\rho$ with temperature only have been observed so far [14]. The "metallic" behavior (d$\rho$/d$T > 0$) takes place from 0.3 to 1.5K and changes to the "insulating" behavior (d$\rho$/d$T < 0$) at higher temperatures. In the "metallic" region at $T \leq 1$K, $\rho$ depends logarithmically on $T$ (Fig. 1b).

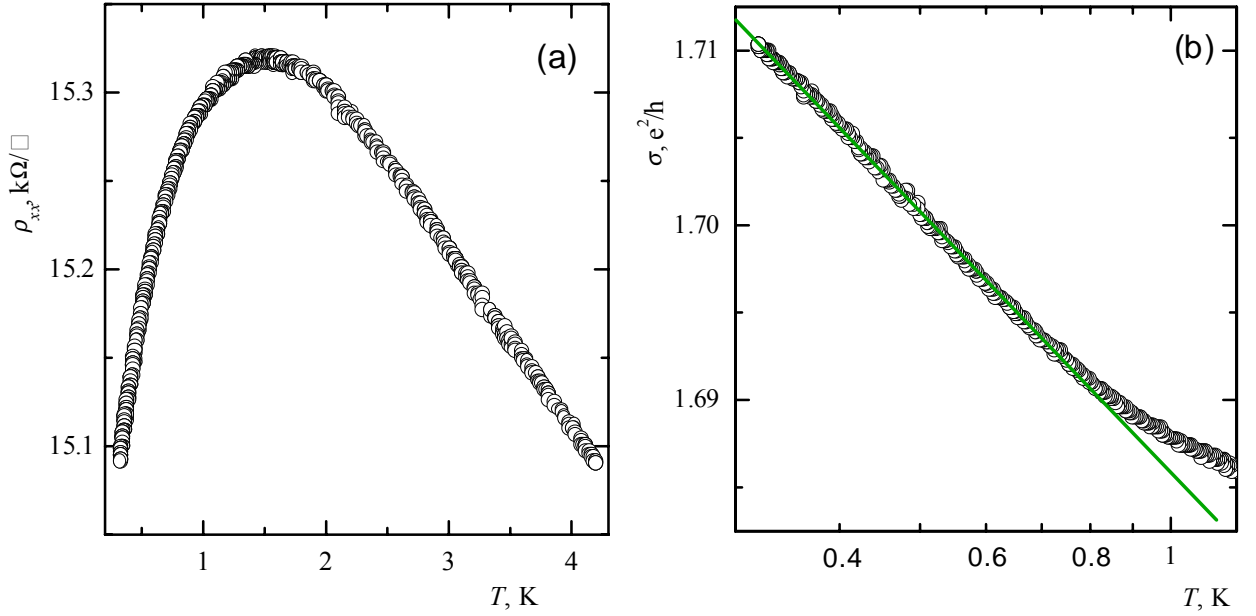

FIG. 1. (a) Temperature dependence of zero-field resistivity.

(b) Zero-field conductivity as a function of ln$T$.

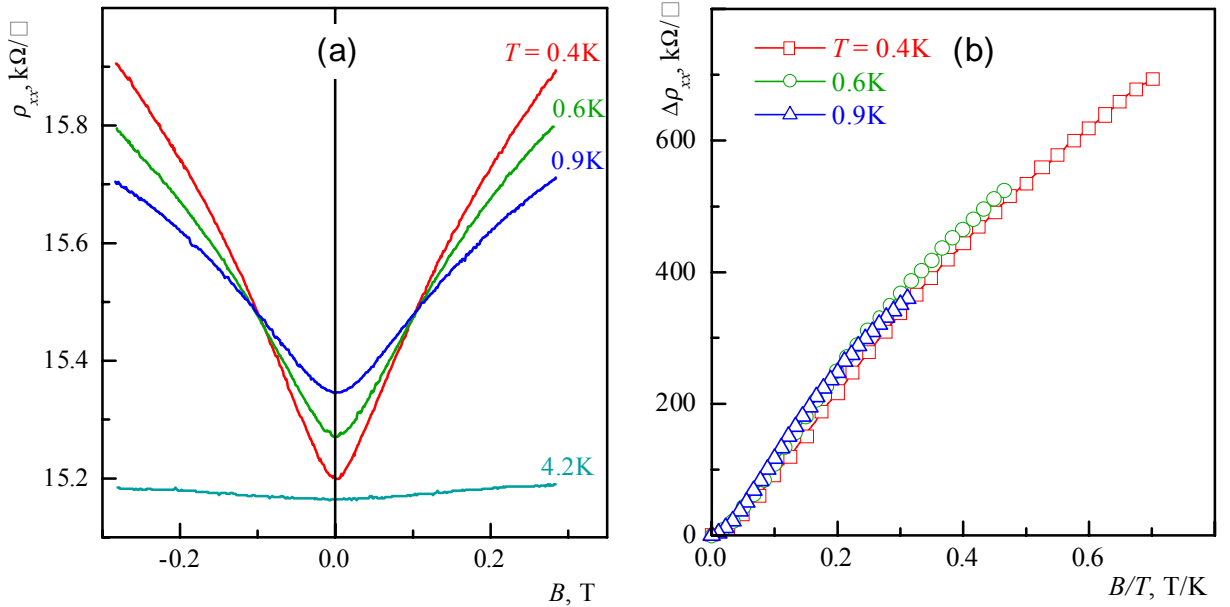

FIG. 2. (a) Dependences of resistivity on perpendicular to plane magnetic field at different temperatures.

(b) Magnetoresistivity as a function of $B/T$.

Fig. 2 shows magnetoresistance data at $T = (0.3 \div 0.9)$K and $T = 4.2$K. Note that *positive* magnetoresistance (PMR) is observed at all temperatures and the upturn of MR is the sharper the lower is temperature (Fig. 2a). For the lowest temperatures $T \leq 0.9$K, the magnetoresistance curves scales as a function of $B/T$ (Fig. 2b).

For investigated samples the parameter $\varepsilon_F \tau/\hbar \cong 1$ and thus formally we are near the critical region of the low-temperature transition from insulating to conducting behavior, which is seen experimentally in a variety of high-mobility semiconductor systems [3-5]. We will take it into account in our analysis of experimental data.

### a) $\rho(B,T)$ dependencies at $T < T_{max}$ (= 1.5K).

It is essential that at $\varepsilon_F \tau/\hbar \cong 1$ the temperature range $kT << \varepsilon_F$ inevitably corresponds to diffusive regime for EEI effect, $kT\tau/\hbar << 1$. Using the Shubnikov de Haas data for effective mass $m/m_0 = 0.08$ ($m_0$ is the free electron mass) we have Fermi energy $\varepsilon_F \cong 3.0$meV and $\hbar/\tau \cong 3.5$meV (the elastic mean free time $\tau \cong 1.9 \times 10^{-13}$s) for investigated sample. Then an estimation gives $kT\tau/\hbar \cong 2.5 \times 10^{-2}$[K$^{-1}$]$\times T$ and, hence, at $T < T_{max}$ (= 1.5K) we are really in a diffusive limit: $kT\tau/\hbar \leq 0.04$.

The observed resistivity dependencies $\rho(B,T)$ may be attributed to the quantum conductivity corrections due to both WL $\delta\sigma_{WL}$ and EEI $\delta\sigma_{ee}$. For the interaction effect in diffusive regime, we have [1, 2]:

$$\delta\sigma_{ee}(B,T) = \delta\sigma_{ee}(T) + \delta\sigma_z(B) \qquad (1)$$

where

$$\delta\sigma_{ee}(T) = \frac{e^2}{2\pi^2\hbar}(1-3\lambda)\ln\frac{kT\tau}{\hbar} \qquad (2)$$

is the zero-field part and

$$\delta\sigma_z(B) = -\frac{e^2}{2\pi^2\hbar}G(B/B_z) \qquad (3)$$

is the Zeeman part of the total EEI correction.

The first term in the factor $(1-3\lambda)$ of Eq. (2) corresponds to the exchange contribution and the second one to the Hartree (triplet) contribution in the zero-field part of $\delta\sigma_{ee}$. The contributions from the exchange and triplet channels have different sign favoring localization or antilocalization, respectively. Here [9, 15]

$$\lambda = \frac{1+\gamma_2}{\gamma_2}\ln(1+\gamma_2) - 1 \qquad (4)$$

where the parameter $\gamma_2$ is the Fermi-liquid amplitude that in a diffusive regime controls EEI in the triplet channel normalized by the density of states.

The function $G(B/B_z)$ in Eq. (3) with $B_z = kT/g\mu_B$ ($g$ is the electron Lande factor and $\mu_B$ is the Bohr magneton) describes the effect of Zeeman splitting on EEI that leads to *positive* MR due to suppression of a great part of antilocalizing triplet contribution into $\delta\sigma_{ee}$. The expression for it was first deduced by Lee and Ramakrishnan for weak EEI ($\gamma_2 << 1$) [16] and then by Castellani, Di Castro and Lee for any value of $\gamma_2$ [17]. At present the $G(B/B_z)$ expression for arbitrary strength of interaction is anew derived as a diffusive limit of the more general formulas [18].

Hole gas in Ge quantum wells for investigated $p$-Ge/Ge$_{1-x}$Si$_x$ heterostructures is described by Luttinger Hamiltonian with effective $g$-factor highly anisotropic relative to mutual orientation of magnetic field and 2D plane: at the bottom of the ground hole subband $g_\perp = 6\kappa \cong 20.4$ (where Luttinger parameter $\kappa \cong 3.4$ for Ge [19]) and $g_\parallel = 0$ for magnetic fields perpendicular ($B_\perp$) and parallel ($B_\parallel$) to 2D plane, respectively [20, 21].

For WL effect at $B = 0$ we have [1, 2]

$$\delta\sigma_{WL}(T) = \frac{e^2}{2\pi^2\hbar} p \ln\frac{T}{T_0}. \tag{5}$$

where $p$ is an exponent in $T$-dependence of phase breaking time, $\tau_\varphi = T^{-p}$. The dependence of $\delta\sigma_{WL}$ on a perpendicular field at $B \ll B_{tr}$, $B_\varphi \ll B_{tr}$ ($B_{tr} = \hbar c/4eD\tau$, $B_\varphi = \hbar c/4eD\tau_\varphi$, where $D$ is diffusion constant) is described by the expression [22]

$$\delta\sigma_{WL}(B) = \frac{e^2}{2\pi^2\hbar}\left[\Psi\left(\frac{1}{2} + \frac{B_\varphi}{B}\right) - \ln\frac{B}{B_\varphi}\right]. \tag{6}$$

For our sample $B_{tr} \cong 1.5$T, so that at $B \leq 0.3$T the inequality $B \ll B_{tr}$ is rather well satisfied. The Eq. (6) leads to a *negative* magnetoresistance (NMR) due to suppression of WL by a magnetic field (dephasing effect). Note that $\delta\sigma_{WL}$ depends on the ratio $B/B_\varphi$ and thus for $p = 1$ [2, 23] it is a function of $B/T$ only.

For resistivity at $B = 0$ we have from (2) and (5) [9, 15]:

$$\frac{d\tilde\rho}{d\xi} = \tilde\rho_0^2(p + 1 - 3\lambda) \tag{7}$$

where $\tilde\rho = (2\pi^2\hbar/e^2)\rho$, $\xi = -\ln(kT\tau/\hbar)$ and $\rho_0 = 1/\sigma_0$. Comparing the observed temperature dependence of $\rho$ in a region of "metallic" conduction at T < 1K (see Fig.1a) with Eq. (7) we conclude that such a behavior is an evidence of predominant role of antilocalizing triplet channel (note that $d\rho/dT > 0$ corresponds to $d\tilde\rho/d\xi < 0$). From fitting with $p = 1$ we have $\lambda = 0.69$ which according to Eq. (4) gives $\gamma_2 = 2.25$ or in designations of Ref. [6]: $F_0^\sigma = -\gamma_2/(1+\gamma_2) = -0.69$.

Next, we find that a dependence of $\rho$ on magnetic field at $T < 1$K, namely, on a ratio $B/T$ (see Fig. 2b), may be quantitatively described only by a combination of both PMR due to Zeeman splitting (3) and NMR due to WL dephasing effect (6) with some predominance of the first one. As an example we present an expression of $\delta\sigma(B) = \delta\sigma_z(B) + \delta\sigma_{WL}(B)$ at weak field limit $B \ll B_z$ [17, 18], $B \ll B_\varphi$ [22]:

$$\delta\sigma(B,T) = \frac{e^2}{2\pi^2\hbar}\left[-0.091\gamma_2(1+\gamma_2) + 0.042\left(\frac{B_z}{B_\varphi}\right)^2\right]\left(\frac{B}{B_z}\right)^2 \tag{8}$$

where (with $p = 1$) a ratio $B_z/B_\varphi$ is $T$-independent.

Really, from the definitions of $B_z$ and $B_\varphi$ we have

$$\frac{B_z}{B_\varphi} = \frac{8\varepsilon_F\tau/\hbar}{g\cdot m/m_0}x, \tag{9}$$

where $x = kT\tau_\varphi/\hbar$. Dephasing time in disordered 2D-system in diffusive limit is determined by self-consistent equation [2, 23, 24]:

$$\Lambda x \ln x = 2\varepsilon_F\tau/\hbar \tag{10}$$

with $\Lambda = 1$ for weak EEI ($\gamma_2 \ll 1$) [2, 23] and

$$\Lambda = 1 + \frac{3\gamma_2^2}{\gamma_2 + 2}$$

at an arbitrary value of $\gamma_2$ [24] (considering the relation $F_0^\sigma = -\gamma_2/(1+\gamma_2)$). The solution of Eq. (10) may be written as

$$x = f(\gamma_2, \varepsilon_F \tau/\hbar) \qquad (11)$$

and for a ratio $B_z/B_\varphi$ we then have

$$\frac{B_z}{B_\varphi} = \frac{8\varepsilon_F \tau/\hbar}{g \cdot m/m_0} \cdot f(\gamma_2, \varepsilon_F \tau/\hbar) \qquad (12)$$

where the right hand part does not depend on temperature.

For magnetoresistance $\Delta\rho_{xx} = \rho_{xx}(B, T) - \rho(0, T)$, we have

$$\Delta\rho_{xx}(B,T)/\rho_0 = -\delta\sigma(B,T)/\sigma_0$$

and a fitting of formulas (3) and (6) to $\Delta\rho_{xx}(B/T)$ dependence in a range of magnetic fields up to 0.3T (see Fig. 2b) gives an opportunity to estimate both the $g$-factor and dephasing time $\tau_\varphi$. In the fitting procedure we use formulas (15-17) of Ref. [18] for $G(B/B_z)$ function in (3), describing Zeeman effect on EEI in diffusive limit. Important is that, according to Ref. [18] [see formulas (9, 10) of that reference], the argument of $G$-function, $B/B_z \equiv g\mu_B B/kT$, depends only on the *bare* electron $g$-factor, which is not renormalized by the Fermi liquid EEI (for a given semiconductor, it is the electron $g$-factor).

We have found that $B_z/B_\varphi \cong 3.7$, $g \cong 14$ ($\pm 1.4$) and $kT\tau_\varphi/\hbar \cong 1$. The obtained value of $g$-factor is somewhat lower than the theoretical result for $\varepsilon_F \to 0$ ($g_\perp = 20.4$) that may be caused by a nonparabolicity of the ground hole subband in the Ge quantum well. The estimation for $\tau_\varphi$ is in rather good accordance with numerical solution of Eq. (10): for $\gamma_2 = 2.25$ and $\varepsilon_F \tau/\hbar = 0.86$ we have $x = 1.33$. The main result of fitting is that we obtain a right order of magnitudes for both the $g$-factor and $\tau_\varphi$.

*b) Magnetic field induced metal-insulator transition.*

In Fig. 3 shown is the resistivity of investigated sample as a function of temperature in several fixed magnetic fields between 0 and 0.3T. It is seen that the effect of $B$ is mainly observed for $T < T_{max}$ where the conducting ("metallic") phase to insulating phase transition takes place at $B \cong 0.1T$. We believe that the transition is induced by Zeeman splitting in the electron spectrum that leads to effective suppression of antilocalizing triplet channel in favor of localizing exchange channel in the total interaction correction $\delta\sigma_{ee}$ [16-18].

The suppression of low-temperature conducting phase by *a parallel* to the 2D-plane magnetic field $B_\parallel$ has been first observed in high-mobility Si-MOSFET for electron densities near the zero-field conductor-insulator transition [11, 25]. $B_\parallel/T$ scaling of the magnetoconductance has been found [12] and such a behavior attributed just to the Zeeman splitting, the $\delta\sigma_z(B_\parallel, T)$ dependence being fitted to the form suggested by [17] (see Eq. (8)) with $\gamma_2 \cong 1.3$. Effect of Zeeman splitting on in-plane magnetoconductivity of high-mobility Si-MOSFET in ballistic regime has been investigated by Pudalov et al. [8] and Vitkalov et al. [26]. For another recent data on parallel magnetic field effect on 2D conducting low-temperature phase see the review papers [5].

In an electron system, the Zeeman splitting effect suppresses a conducting phase independently on the angle between the field and the 2D-plane (see, for example, [25]). But for a *hole* system with highly anisotropic $g$-factor ($g_\parallel \ll g_\perp$) this effect in parallel magnetic field should be weakened considerably. Thus, in the *perpendicular* to the 2D-plane magnetic field we investigate a situation where a WL effect should be present equally with interaction correction, and the usual negative WL magnetoresistance should be observed. Really, the high-mobility electron Si-MOSFET structures exhibit the weak negative magnetoresistance attributed to the orbital single-particle quantum interference correction (i.e. to WL effect) in the *low perpendicular* magnetic field $B_\perp < 0.1$T [25].

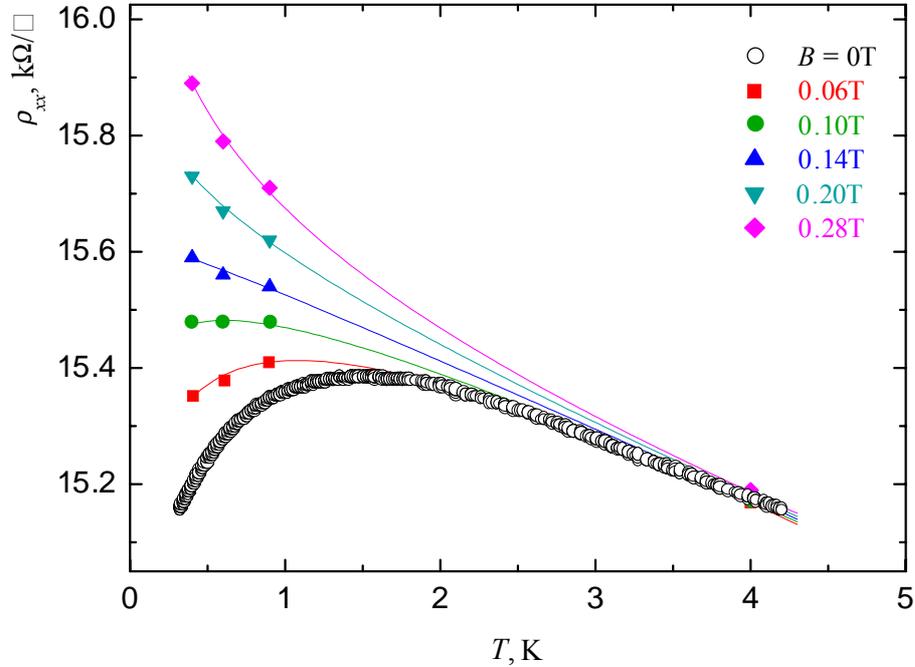

FIG. 3. Magnetoresistivity as a function of ln$T$ in different magnetic fields.

Magnetoresistance in *perpendicular* fields for *p*-SiGe samples on the metallic side of the $B = 0$ metal – insulator transition has been investigated by Coleridge et al. [13]. Magnetoresistance shows clear evidence of both quantum interference and Zeeman interaction effects. The initial NMR attributed to dephasing by the magnetic field due to the WL term is followed by a PMR due to the term identified with the Zeeman interaction effects. The Zeeman term which scales as $B/T$ could not be quantitatively described by a conventional theory for a weakly interacting 2D-system [16]. The best fit of the data has been obtained using low- and high-field limits of Castellany et al. theory [17] with the value of $\gamma_2$ up to 2.6.

In a recent work of Gao et al. [27] on the *p*-GaAs system that is metallic at $T \leq 0.3$K, only a negative low-field MR in perpendicular fields shows up, so that the WL effect overwhelms the effect of Zeeman splitting observed in *p*-GaAs in parallel magnetic field [28, 29].

In contrast to electron Si-MOSFET system [25] or hole SiGe [13] and GaAs [27] systems we do not observe the low-field NMR and believe that this is a consequence of a particular parameter relation characterizing the investigated *p*-Ge quantum wells, specifically, of a large value of the hole band *g*-factor (see Eqs. (8) and (12)).

*c) Nonmonotonic temperature dependence of resistivity at B = 0.*

There exist at least two approaches in explanation of the nonmonotonic temperature dependence of resistivity with low temperature "metallic" phase: for pure diffusive [9] and pure ballistic [10] regimes (see review papers [5]).

In the paper of Punnoose and Finkelstein [9] the highly nonmonotonic temperature dependence of resistivity in a Si-MOSFET sample close to the critical region of the metal-insulator transition ($\rho(T_{max}) \leq h/e^2$) has been well described on a basis of the renormalization group (RG) theory [15, 17]. According to [15, 17] the interplay of EEI and disorder leads to such a renormalization of Fermi-liquid parameter $\gamma_2$ that it increases monotonically as the temperature is lowered. When $\gamma_2$ increases beyond the value $\gamma_2^*$, for which $(p + 1 - 3\lambda) = 0$ (see Eg.(7)), the resistivity passes through a maximum.

In the presence of two valleys in Si-MOSFET the increase in the number of multiplet channels for EEI from 3 to 15 takes place and $\gamma_2^*$ reduces considerably from $\gamma_2^* = 2.04$ for $n_v = 1$ (as in our work) to $\gamma_2^* = 0.45$ for $n_v = 2$. This strong reduction of $\gamma_2^*$ makes it possible to reach a resistivity maximum at not exponentially small temperatures. In contrast, the large value $\gamma_2^* = 2.04$ for a single valley makes it difficult for the nonmonotony to be observed. Moreover, it is emphasized in Ref. [9] that such a scaling behavior is to be realized only in ultraclean samples which is really not the case for our structures.

On the other hand, Das Sarma and Hwang [10] have explained a transition from "metallic" to apparent "insulating" phase with increasing $T$ on the basis of quasiclassical theory of temperature dependent screening of impurity potential [30, 31], suggesting that the resistivity maximum is due to a crossover between the Fermi and Boltzmann statistics. In Ref. [6, 7] it is argued that this approach has a common physical origin with the EEI effect at $kT\tau/\hbar \gg 1$, i.e. in a limit of single-impurity scattering, and that the theory of EEI correction in ballistic regime provides a systematic microscopic justification of the concept of temperature dependent screening (see Sec. IIIF of Ref. [6] and Sec. IV of Ref. [7]).

We speculate that in our experiment the crossover between diffusive and ballistic regimes with an assumption of smooth character of random impurity potential may be responsible for the nonmonotonic $\rho(T)$ dependence. Really, in a smooth disorder (small angle scattering) the EEI contribution in ballistic regime, which is proportional to the return probability after a single-scattering event, vanishes as $\exp(-k_F d)$, with $d$ being a spatial range of random impurity potential [7]. As shown in [6, 7], the crossover between diffusive and ballistic limits should take place at small values of $kT\tau/\hbar \cong 0.1$ since the natural dimensionless variable of the theory is $2\pi kT\tau/\hbar$. A flat region in $\rho(B)$ dependencies of a high-mobility n-GaAs heterostructure at $T \geq 1.2$K in fields $\omega_c \tau < 1$ (after the initial rapid drop of MR due to WL dephasing) the authors of [32] interpret as a clear indication that in the ballistic regime the long-range potential suppresses the zero-field interaction correction.

For a range of impurity potential in our structures we have $k_F d \cong 1$, where $d \cong 100$Å is an effective spacer width [20, 33]. Thus, we believe that just a suppression of EEI correction with $T$-increasing due to a gradual change of regime results in a transition to the "insulating" behavior at $T > T_{max}$ where WL effect becomes predominant. Note that for a point-like scatterers the linear-in-$T$ contribution

$$\delta\sigma = \frac{e^2}{2\pi^2\hbar}\left(1 + \frac{3F_0^\sigma}{1+F_0^\sigma}\right)\frac{kT\tau}{\hbar} \equiv \frac{e^2}{2\pi^2\hbar}(1-3\gamma_2)\frac{kT\tau}{\hbar}$$

will come to light in total $\delta\sigma_{ee}$ correction with a transition to ballistic regime [6]. For our value of $\gamma_2$ we have $(1-3\gamma_2) \cong -5.75$ and this contribution should lead to a steeper increase of $\rho(T)$ ("more metallic" behavior) with $T$ increasing, which obviously is not the case in our experiment.

## 3. Conclusions

We think that a mutual compensation of WL and EEI effects takes place for investigated p-Ge/Ge$_{1-x}$Si$_x$ heterostructure with parameter values near a nominal 2D "metal-insulator transition", $\rho \cong h/e^2$ ($\varepsilon_F\tau/\hbar \cong 1$). In the pure diffusive regime $kT\tau/\hbar < 0.025$ ($T < 1$K), the predominance of the antilocalizing triplet channel contribution into EEI correction leads to an apparent metallic behavior, $d\rho/dT > 0$. But with a crossover to the ballistic regime (at $kT\tau/\hbar \cong 0.1$) the gradual reduction of EEI contribution in favor of WL one ($d\rho/dT < 0$) occurs for a smooth or predominantly smooth disorder.

Due to the high value of the Ge valence band *g*-factor, the effect of Zeeman splitting in perpendicular to 2D plane magnetic field causes an effective suppression of the triplet channel contribution and conduces to the insulating behavior of $\rho(T)$ in the whole temperature interval.

Finally, we compare our results for Fermi-liquid amplitude $\gamma_2$ ($F_0^\sigma$) with those obtained from an analysis of experimental data on other semiconductor systems in *diffusive* regime. The highest values of the parameter $\gamma_2$ reported for low-mobility [34, 35] and high-mobility [12] Si-MOSFET as well as for *p*-SiGe [13] and *p*-Ge (this work) are presented in the Table. Here $r_s = (\pi n)^{-1/2}/a_B$ is the usual dimensionless Wigner-Seitz interaction parameter with *n* as the density of carriers and $a_B$ as the effective semiconductor Bohr radius. Let us note that all the values of $F_0^\sigma$ shown in the table are appreciably larger in magnitude (for similar values of $r_s$) that those obtained from an analysis of transport effects in terms of recent EEI theories in ballistic regime (see Ref. [36]).

Table. Some experimental data for Fermi-liquid interaction parameter in diffusive regime.

| Semiconductor | $\varepsilon_F\tau/\hbar$ | $r_s$ | $\gamma_2$ | $F_0^\sigma$ | Reference |
|---|---|---|---|---|---|
| Si-MOSFET | 1.25 | – | 3.5 | -0.78 | [34] |
| Si-MOSFET | 1.3 | 1.6 | 3.2 | -0.76 | [35] |
| *p*-SiGe | 7.2 | 4 | 2.6 | -0.72 | [13] |
| *p*-Ge | 0.86 | 1.75 | 2.25 | -0.69 | This work |
| Si-MOSFET | 0.93 | 5.6 | 1.3 | -0.56 | [12] |

It is also seen that almost all of the data in the table (with an exception of the result of Ref. [13]) correspond to a region of nominal metal-insulator transition with $\varepsilon_F\tau/\hbar \cong 1$. Then it may be that such a high $\gamma_2$ value is a consequence of the renormalization of the Fermi-liquid parameter due to an interplay of interaction and disorder in the diffusive regime, which, according to RG theory [9, 15, 17], is especially significant just in a proximity of $\rho = h/e^2$, i.e. for $\varepsilon_F\tau/\hbar \cong 1$. On the other hand, an apparent reducing of the interaction amplitude extracted from the temperature dependence of resistivity in the ballistic regime may be related to a mixed (point-like plus smooth) character of the random impurity potential (see Eq.(2.53) of Ref.[7]).

## Acknowledgements

We are grateful to P.T. Coleridge for turning our attention to significance of Zeeman effect in a hole system and to V.I. Okulov and S.G. Novokshonov for helpful discussions.
The work is supported by Russian Foundation for Basic Researches, projects 04-02-16614, 02-02-16401 and the RAS program "Physics of solid state nanostructures".
.

## References

1. P.A. Lee, T.V. Ramakrishnan, Rev. Mod. Phys. **57**, 287 (1985).
2. B.L. Altshuler, A.G. Aronov in "Electron-electron interactions in disordered systems", ed. by A.L.Efros and M.Pollak (Elsevier, Amsterdam 1985) p.1.
3. S.V. Kravchenko, G.V. Kravchenko, J.E. Furneaux, V.M. Pudalov, and M. D'Iorio, Phys. Rev. B **50**, 8039 (1994).
4. B.L. Altshuler, D.L. Maslov, and V.M. Pudalov, Physica E **9**(2), 209 (2001).
5. E. Abrahams, S. V. Kravchenko, and M.P. Sarachik, Rev. Mod. Phys. **73**, 251 (2001); S.V. Kravchenko, M.P. Sarachik, Rep. Prog. Phys. **67**, 1 (2004)
6. G. Zala, B.N. Narozhny and I.L. Aleiner, Phys. Rev. B **64**, 214204 (2001); G. Zala, B.N. Narozhny, and I.L. Aleiner, cond-mat/0107333
7. I.V. Gornyi and A.D. Mirlin, Phys. Rev. Lett. **90**, 076801 (2003); I V. Gornyi and A.D. Mirlin, cond-mat/0306029.
8. V.M. Pudalov, M.E. Gershenson, H. Kojima, G. Brunthaler, A. Prinz and G. Bauer, Phys. Rev. Lett. **91**, 126403 (2003); cond-mat/0401031.
9. A. Punnoose, A.M. Finkelstein, Phys. Rev. Lett. **88**, 016802 (2002).
10. S. Das Sarma and E. H. Hwang, cond-mat/0302047, cond-mat/0302112.
11. D. Simonian, S.V. Kravchenko, M.P. Sarachik and V.M. Pudalov, Phys. Rev. Lett. **79**, 2304 (1997).
12. D. Simonian, S.V. Kravchenko, M.P. Sarachik and V.M. Pudalov, Phys. Rev. B **57**, R9420 (1998).
13. P.T. Coleridge, A.S. Sachrajda and P.Zawadzki, cond-mat/9912041, cond-mat/0011067, Phys. Rev. **B65**, 125328 (2002).
14. Yu.G. Arapov, V.N. Neverov, G.I. Harus, N.G. Shelushinina, M.V. Yakunin, O.A. Kuznetsov, Semiconductors **32** 649 (1998); cond-mat/0203435; cond-mat/0212612.
15. A.M. Finkelstein, Zh. Eksp. Teor. Fiz. **84**, 168 (1983) [Sov. Phys. JETP **57**, 97 (1983)]; A.M. Finkelstein, Z. Phys. B **56**, 189 (1984).
16. P.A. Lee, T.V. Ramakrishnan, Phys. Rev. B **26**, 4009 (1982).
17. C. Castellani, C. Di Castro, P.A.Lee, Phys. Rev. B **57**, R9381 (1998).
18. G. Zala, B.N. Narozhny, and I.L. Aleiner, Phys. Rev. B **65**, 020201 (2001).
19. J.C. Hensel, K. Suzuki, Phys. Rev. Lett. **22**, 838 (1969).
20. Yu.G. Arapov, O.A. Kuznetsov, V.N. Neverov, G.I. Harus, N.G. Shelushinina, M.V. Yakunin, Semiconductors, **36**, 519 (2002).
21. A.V. Nenashev, A.V. Dvurechenskii, A.F. Zinov'eva, Zh. Eksp. Teor. Fiz., **123**, 362 (2003) [JETP **96**, 321 (2003)].
22. S. Hikami, A.I. Larkin, I. Nagaoka, Progr. Teor. Phys. **63**, 707 (1980).
23. I.L. Aleiner, B.L. Altshuler, M.E. Gershenson, cond-mat/9808053.
24. B.N. Narozhny, G. Zala, I.L. Aleiner, Phys, Rev. B, **65**, 180202R (2002).
25. V.M. Pudalov, G. Brunthaler, A. Prinz, and G. Bauer, Pis'ma v ZhETF **65**, 887 (1997) [JETP Lett. **65**, 932 (1997)].
26. S.A. Vitkalov, K. James, B.N. Narozhny, M.P. Sarachik, and T.M. Klapwijk, cond-mat/0204566.
27. X.P.A. Gao, A.P. Mills Jr., A.P. Ramirez, L.N. Pfeifer, K.W. West, cond-mat/0308003.
28. J. Yoon, C.C. Li, D. Shahar, D.S. Tsui, and M. Shayegan, cond-mat/9907128.
29. H. Noh, M.P. Lilly, D.C. Tsui, J.A. Simmons, E.H. Hwang, S. Das Sarma, L.N. Pfeiffer, and K.W. West, Phys. Rev. **B68**, 165308 (2003).
30. F. Stern, Phys. Rev. Lett., **44**, 1469 (1980); F. Stern, S. Das Sarma, Solis State Electron., **28**, 158 (1985); S. Das Sarma, Phys. Rev. B, **33**, 5401 (1986).
31. A. Gold, V.T. Dolgopolov, Phys. Rev. B, **33**, 1076 (1986).
32. L. Li, Y.Y. Proskuryakov, A.K. Savchenko, E.H. Linfield, D.A. Ritchie, cond-mat/0207662.

33. Yu.G. Arapov, G.A. Alshanskii, G.I. Harus, V.N. Neverov, N.G. Shelushinina, M.V. Yakunin and O.A. Kuznetsov, Nanotechnology, **13**, 86 (2002).
34. M.S. Burdis and C.C. Dean, Phys. Rev. B, **38**, 3269 (1988).
35. D.J Bishop, R.C. Dynes, and D.C. Tsui, Phys. Rev. B, **26**, 773 (1982).
36. E.A. Galaktionov, A.K. Savchenko, S.S. Safonov, Y.Y. Proskuryakov, L. Li, M. Pepper, M.Y. Simmons, D.A. Ritchie, E.H. Linfield, and Z.D. Kvon, cond-mat/0402139.